\documentclass[a4paper,11pt]{article}
\usepackage{amssymb,amsmath,bm,latexsym,color,epsf,cite,comment}
\usepackage{graphicx}
\usepackage{braket}
\usepackage{xcolor}

\newcommand{\bea}{\begin{eqnarray}}
\newcommand{\eea}{\end{eqnarray}}
\newcommand{\vs}[1]{\vspace{#1 mm}}

\renewcommand{\a}{\alpha}
\renewcommand{\b}{\beta}

\renewcommand{\d}{\delta}

\newcommand{\s}{\sigma}

\newcommand{\tr}{{\rm tr}}
\newcommand{\Tr}{{\rm Tr}}

\def\R{{\mathbb R}} \def\C{{\mathbb C}} 
 \def\one{\mbox{1 \kern-.59em {\rm l}}}

\def\cM{{\mathcal M}} 
\def\cH{{\mathcal H}}

\begin{document}

\begin{flushright}
 UWThPh 2025-17 \\
 NITEP 258

\end{flushright}

\begin{center}
{\Large\bf General Relativity in IIB Matrix Model}
\vs{10}

{\large
Pei-Ming Ho$^{a,b,}$\footnote{e-mail address: pmho@phys.ntu.edu.tw}, 
Hikaru Kawai$^{a,b,c,}$\footnote{e-mail address: hikarukawai@phys.ntu.edu.tw},
Harold C. Steinacker$^{d,}$\footnote{e-mail address: harold.steinacker@univie.ac.at}
} \\
\vs{5}

$^a${\em Department of Physics and Center for Theoretical Physics, National Taiwan University, Taipei 106, Taiwan}

$^b${\em Physics Division, National Center for Theoretical Sciences, Taipei 106, Taiwan}
\vs{2}

$^c${\em Nambu Yoichiro Institute of Theoretical and Experimental Physics (NITEP), Osaka Metropolitan University, Osaka 558-8585, Japan}

$^d${\em Faculty of Physics, University of Vienna Boltzmanngasse 5, A-1090 Vienna, Austria}

\vs{5}
{\bf Abstract}
\end{center}

The matrix models are non-perturbative formulations of string theory, from which many believe that spacetime arises. The matrix fluctuations around the spacetime thus created should represent both matter and gravitational fields. 
In this paper, we discuss how the gravitational field emerges from the IIB matrix model. 
In particular, we consider how diffeomorphism invariance arises and how unitarity is guaranteed in this theory.
Specifically, we consider matrices as bilocal fields and discuss how the Lorentz-invariant vacuum and low-energy excitations around it can be expressed. We then discuss how the conditions for the theory to be unitary can be written in terms of bilocal fields. We argue that in the low-energy limit, the bilocal fields are reduced to local fields consisting of a finite number of massless fields and an infinite number of massive fields, satisfying unitarity.


\vs{10}

\setcounter{footnote}{0}


\section{Introduction} \label{int}

The matrix models are a non-perturbative formulation of string theory, and space-time is thought to emerge as the configuration of the matrices \cite{BFSS, IKKT}.
Considering quantum fluctuations around the emergent spacetime, it has been confirmed in various cases that gravity is induced.

In particular, in the case of the IIB matrix model \cite{IKKT}, the emergence of spacetime can be roughly classified into three cases. First, if we consider the IIB matrix model as a matrix regularization of the worldsheet of string theory, the matrices represent the spacetime coordinates (and their supersymmetric partners). On the other hand, when the IIB matrix model is regarded as a large-N reduced model \cite{EK}, the diagonal elements of the matrices represent the spacetime momenta. 
In this case, there are two further cases.
The first is the case of the quenched reduced model \cite{Parisi, BHN, GK}, and the second is the case of the twisted reduced model \cite{GO, GKA}. 
The former case represents a commutative spacetime, while the latter case represents a non-commutative spacetime.
In the end, the emergence of spacetime in the IIB matrix model can be broadly classified into the following three categories.
\begin{enumerate}
\item  Matrices represent spacetime coordinates.
\item  Matrices represent momenta.
\item  Matrices represent noncommutative spacetime.
\end{enumerate}
Although in basic noncommutative spaces, such as $\R^{1,3}_\theta$, coordinates and momenta are not distinguished, one must make this distinction when considering covariant quantum spacetimes \cite{Sperling:2019xar,Steinacker:2017vqw}.

Due to the large symmetry of $\text{SU}(\infty)$, 
these categories may eventually become equivalent, but the details are currently unknown. Let us summarize how the appearance of gravity is currently confirmed in each case.


In the first category, a one-loop calculation of the force between two classical D-strings yields a force due to the exchange of graviton and Kalb-Ramond fields \cite{IKKT}, reproducing the result for type IIB string theory. It has recently been shown that the corresponding matrix model has a partition function that matches type IIB supergravity in all orders when $1/2$ supersymmetry is preserved by introducing an appropriate background to the Kalb-Ramond field \cite{HL, Ko}.
As an alternative approach, it can be shown that the loop equations reproduce the string field theory under several assumptions, but so far these assumptions have not been proven \cite{AIKKTT}.

The discussion of the emergence of general relativity within the second category is the central issue of this paper. So far, this does not appear to have been very successful \cite{HKK}. However, the purpose of this paper is to advance the discussion by analyzing the matrix model in terms of bilocal fields. Indeed, it can be shown that gravity appears universally in the low-energy effective theory, i.e., independent of the details of the model.

As for the third category, gravity has also been studied intensively as a field theory on noncommutative spaces. 
In \cite{St}, it was shown that fluctuations around basic noncommutative backgrounds of the IIB matrix model are related to gravity. 
The Einstein-Hilbert action was shown to arise at one loop \cite{Steinacker:2023myp}, notably for covariant quantum spacetimes \cite{Sperling:2019xar} which lead to a higher-spin extension of gravity.

This paper is organized as follows.
First, in section \ref{mat}, we discuss that the matrix can be regarded as a bilocal field. 
In section \ref{flu}, we study excitations around backgrounds consisting of commuting matrices and summarize the quenched large-$N$ reduced model; in section \ref{com}, we investigate how such a commutative background is expressed in terms of bi-local fields and discuss how the vacuum is specified by the expectation value of the bi-local fields. 
We also show that the bilocal field naturally satisfies positive definiteness or unitarity.
In section \ref{gra}, we apply the discussion of the possibility of gravitons appearing as Nambu-Goldstone bosons to the IIB matrix model and discuss the emergence of gravity automatically in the low-energy effective theory.
In section \ref{low}, we introduce a model of low-energy effective theory in which matrices are represented by divergence-free vector fields, i.e., the minimal model. 
In section \ref{constraint}, we propose a general condition that bilocal fields must satisfy in order for the theory to be unitary. In fact, this condition is so natural that the minimal model is derived in the low-energy limit.
In section \ref{phy}, we examine the physical states of the minimal model. We also discuss the difficulties involved in representing matrices in terms of differential operators. 
In section \ref{sec:NC-backgrounds}, we explain how these issues are resolved on noncommutative backgrounds, and clarify the relation and distinction from the commutative case.
Section \ref{sum} is a summary.


\section{Matrix as bi-local field} \label{mat}

It is widely accepted that matrix models are strong candidates for non-perturbative formulations of string theory. 
In these models, spacetime is generated from matrices, and the fluctuations of the matrices around the spacetime thus generated should represent both matter and gravitational fields.
In the following, we will discuss how general relativity arises from the IIB matrix model from as general a standpoint as possible. In particular, we will discuss how general coordinate transformation invariance arises from the symmetry of the matrix model and how unitarity is guaranteed in this theory.

We start from the IIB matrix model \cite{IKKT}, the action of which is given by
\bea
{S_{IIB}= \Tr\left(-\frac{1}{4} [A_\mu , A_\nu ]^{2} -\frac{1}{2} \bar{\psi}\Gamma^{\mu}[A_\mu, \psi]\right),}
\label{IKKT}
\eea
where $A_\mu$ and $\psi_s$ are Hermitian operators on a vector space $V$. (The bosonic variables $A_\mu$ are the components of a ten-dimensional vector, and the fermionic variables $\psi_s$ are the components of a Majorana-Weyl spinor in ten dimensions. )

If we take $\mathbb{C}^N$ as $V$, $A_\mu$ and $\psi_s$ are matrices.
\bea
\bra{i}A_\mu \ket{j}=(A_\mu)_{i,j}, \ \bra{i}\psi_s\ket{j}= (\psi_s)_{i,j} .
\label{RepMat}
\eea
This is nothing but the original interpretation.

On the other hand, if we take the space of functions on $\mathbb{R}^d$ as $V$,
\bea
V= \{\phi(x) :\mathbb{R}^d \to \mathbb{C}\},
\label{posSpa}
\eea
$A_\mu$ and $\psi_s$ are bi-local fields.
\bea
\bra{x}A_\mu \ket{y}=A_\mu(x,y), \ \bra{x}\psi_s\ket{y}= \psi_s(x,y) .
\label{bilFie}
\eea
In this case the action becomes
\bea
S=-\frac{1}{2}\int d^d x \ d^d y \ d^d z\  d^d w \ 
\{ A_\mu(x,y)A_\nu(y,z)A_\mu(z,w)A_\nu(w,x) & \notag \\
-A_\mu(x,y)A_\nu(y,z)A_\nu(z,w)A_\mu(w,x) \} \notag \\
-\frac{1}{2}\int d^d x \ d^d y \ d^d z\ 
\{ \bar{\psi}(x,y)\Gamma_\mu A_\mu(y,z)\psi(z,x) & \notag \\
-\bar{\psi}(x,y)\Gamma_\mu \psi(y,z)A_\mu(z,x)\} .
\label{bilAct}
\eea

The above two pictures could be just different ways of choosing the basis of $V$ if $V$ is infinite-dimensional. 
They are. However, the second picture is very useful when discussing how fluctuations, especially gravity, appear around the d-dimensional spacetime emerging from the matrix model. 
From this perspective, it is important to note that the action \eqref{bilAct} is invariant under the diffeomorphisms on $\mathbb{R}^d$:
\bea
&A_{\mu}(x,y) \mapsto \left(J(x)\right)^{\frac{1}{2}} \left(J(y)\right)^{\frac{1}{2}} A_{\mu}(\varphi(x),\varphi(y)), \nonumber \\
&\psi(x,y) \mapsto \left(J(x)\right)^{\frac{1}{2}} \left(J(y)\right)^{\frac{1}{2}} \psi(\varphi(x),\varphi(y)), 
\label{Diff}
\eea
where $\varphi$ is a diffeomorphism on $\mathbb{R}^d$ and
$J(x)$ is its Jacobian
\bea
J(x)=\left|\det(\partial_\alpha \varphi^\beta(x))\right|.
\label{Jacobian}
\eea
In other words, under diffeomorphisms, 
the bi-local fields $A_\mu(x,y)$ and $\psi(x,y)$ transform as square-root densities in both arguments $x$ and $y$.

This symmetry is a small subset of the original invariance under unitary transformations.
In fact, the transformation \eqref{Diff} is given by the unitary transformation
\bea
\ket{x} \mapsto  \left(J(x)\right)^{\frac{1}{2}} \ket{\varphi(x)}.
\label{Diffx}
\eea
In the case of $V=\mathbb{C}^N$, this is nothing but the permutation symmetry of the basis vectors $\ket{i}$.
However, as we will discuss, it plays an important role in the emergence of general relativity around the emergent spacetime. 

As we will see in sections \ref{low} and \ref{constraint} below, it is natural to require the following constraints on the bilocal fields in order to guarantee the unitarity of the theory.
\bea
\int d^d y \, \phi(x,y)=0,
\label{unitarityConstraint}
\eea
where $\phi(x,y)$ is $A_\mu(x,y)$ or $\psi(x,y)$ or, in general, a Hermitian bilocal field.
In other words, instead of considering all Hermitian bilocal fields as the space of matrices, we consider the space with the constraint as in \eqref{unitarityConstraint}.
As discussed in section \ref{constraint}, such a constraint can be written as \eqref{matrix-constraint} and is consistent with the equations of motion.

Then the symmetry of the theory is reduced to the volume preserving diffeomorphisms (VPD) on $\mathbb{R}^d$:
\bea
\phi(x,y) \mapsto \phi(\varphi(x),\varphi(y)), \quad \left|\det(\partial_\mu\varphi^\nu)\right|=1.
\label{VPD}
\eea
This is because the constraint \eqref{unitarityConstraint} is invariant under \eqref{Diff} only when $J = 1$.
The set of VPDs seems a bit too small compared to the full set of diffeomorphisms that would guarantee the emergence of general relativity as a low-energy effective theory.
However, in a variety of cases, invariance under the VPD is often sufficient. 
A typical example is the unimodular gravity in which the determinant of the metric tensor is fixed to one. 
In this case, the symmetry is restricted to the VPD, but as is well known, this system is equivalent to ordinary general relativity. 
In the following, we will discuss that this is indeed the case for the IIB matrix model and show that general relativity arises naturally in the low-energy region.


\section{Fluctuations around commuting background} \label{flu}

In the case $V=\mathbb{C}^N$, one of the simplest classical solutions is given by the diagonal matrices:
\bea
\begin{aligned}
&A_\mu^{(0)}=
\left\{
\begin{alignedat}{2}
&P^{(0)}_\mu \,\,\  (\mu=1,\cdots, d), \quad (P^{(0)}_\mu)_{i,j}=
p_\mu^{i} \ \delta_{i,j} \ ,\\
&\ 0 \quad \left(\mu=d+1 ,\cdots ,10 \right),
\end{alignedat}
\right. \\
&\psi_s^{(0)}=0.
\end{aligned}
\label{diag}
\eea
Here $P^{(0)}_\mu$ are diagonal matrices and the diagonal elements of each matrix are 
$p_\mu^{i}$.

In order to examine the fluctuations around such classical solutions, it is common practice to separate each matrix into its diagonal and off-diagonal components.
\bea
A_{\mu}&=\,P_\mu+\tilde{A}_\mu, \nonumber \\
\psi_s&=\,\theta_s+\tilde{\psi}_s,
\label{dec}
\eea
where $P_\mu$ and $\theta_\mu$ are diagonal matrices and $\tilde{A}_\mu$ and $\tilde{\psi}_\s$ are off-diagonal matrices. Then, by performing path integrals on the off-diagonal elements, the effective action on the diagonal elements can be obtained 
\footnote{Strictly speaking, integrating off-diagonal elements requires fixing the gauge, such as the temporal gauge $\tilde{A}_0=0$ or the Landau gauge $\sum_\mu[P_\mu, A_\mu]=0$.}
\cite{AIKKT}. 
The resulting effective action is complicated, and it is not easy to solve for the exact distribution of the diagonal elements. However, certain approximations, or numerical results for relatively small matrix sizes, suggest that the elements of $P_\mu$ have a four-dimensional distribution \cite{AIKKT, AABHN, KNT}.
Following these analyses, we assume that the matrices fluctuate around the diagonal matrices as in  \eqref{diag}.

In short, the picture we adopt here is as follows. The matrices fluctuate approximately around the diagonal matrices. However, the diagonal and off-diagonal elements interact in complex ways, and the distribution of the diagonal elements is determined as a result of this interaction. In this sense, the spacetime is roughly determined by the distribution of the diagonal elements, and the fluctuations of the matrices around such a configuration should represent the excitations of the matter and gravitational fields in that spacetime.

As a first step to investigate the overall structure of the fluctuations of the matrices, let us examine the situation when the diagonal elements are fixed by hand and only the off-diagonal elements are assumed to be dynamical.
A typical result in that case is known as the quenched large-N reduced model \cite{EK, Parisi, BHN, GK}. 
More precisely, consider the following situations:
\begin{enumerate}
\item When we consider $p_\mu^{i}$ $(i=1,\cdots,N)$ 
as $N$ points in $\mathbb{R}^d$, those points are uniformly distributed in $\mathbb{R}^d$. 
\item The diagonal elements of the matrices are quenched. More precisely, the diagonal elements of $A_\mu$ and $\psi_s$ are not dynamical, but are fixed to those of the matrices given in ~\eqref{diag}.
\item Take the large-$N$ limit. In other words, consider the sum of the planar diagrams.
\end{enumerate}
Then the theory is equivalent to the $d$-dimensional field theory obtained by dimensional reduction from the 10-dimensional super Yang-Mills theory to $d$ dimensions. 

For example, the $L$-point Green's function of $A_\mu$ are given by
\bea
\langle (A_{\mu_1})_{i_1 j_1} \cdots (A_{\mu_L})_{i_L j_L} \rangle_{MM}= \sum_{\text{permutations}}\delta_{j_1i_2}\cdots\delta_{j_L i_1}
\langle\tilde{A}_{\mu_1}(p^{(1)}) \cdots \tilde{A}_{\mu_L}(p^{(L)})  \rangle_{FT},
\label{Lpt}
\eea
where the left-hand side is the connected Green's function of the matrix model. The symbol $\langle\  \rangle_{FT}$ on the right-hand side is the connected Green's function of the large-$N$ field theory for the Fourier transform of the gauge field $A_\mu^a$ with the color factor $\tr(t^{a_1}\cdots t^{a_L})$ removed. The momentum $p^n$ of each field $\tilde{A}_{\mu_n}^{a_n}$ is given by
\bea
p_\mu^{(n)}=p_\mu^{i_n}-p_\mu^{j_n}.
\label{pk}
\eea
The summation on the right-hand side is taken over the permutations of $j_1,\cdots, j_L$.

\begin{figure}[htbp]
  \centering
\vspace{-10mm}
  \includegraphics[width=16cm]{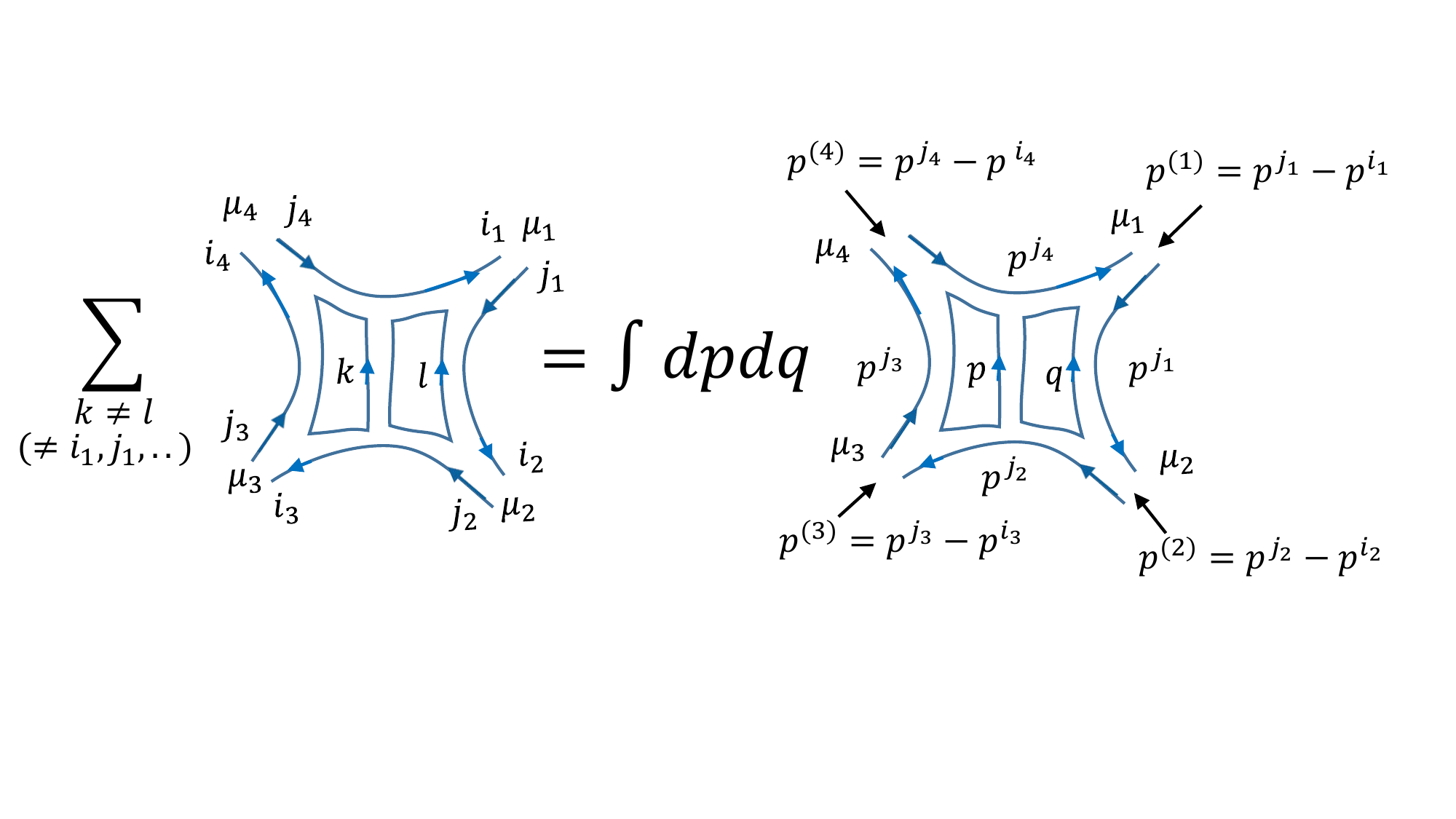}
\vspace{-25mm}
  \caption{The quenched large-N reduced model.}
\label{FigQR}
\end{figure}

This is illustrated in Figure \ref{FigQR}. The left side of the figure shows the case where, among the possible permutations, the indices of the matrices are contracted as $\delta_{j_ii_2} \delta_{j_2i_3}\delta_{j_3i_4}\delta_{j_4i_1}$. 
The crucial point is that the sum over the internal index loops in the matrix model exactly reproduces the integral over the loop momenta in the field theory if the diagram is planar and the distribution of $p_\mu^{(0)\,i}$ is uniform in $\mathbb{R}^d$. Another important point here is that on the matrix model side, the diagonal elements of the matrices are considered quenched, not dynamical. 
That is, only the off-diagonal elements have propagators. 
Therefore, the indices facing each other on the double line must have different values.

In summary, as long as the diagonal elements in the matrix model are fixed to a uniform distribution in $d$-dimensions and only the planar diagrams are considered, the theory is equivalent to the large $N$ gauge theory in $d$-dimensions. 
In other words, if gravity emerges from the matrix model as expected, the degrees of freedom of gravity should arise from the non-planar diagrams and the degrees of freedom of the diagonal elements. 
In the following, we show that this is indeed the case.


\section{Commuting background in terms of bi-local field} \label{com}

To see the degrees of freedom of gravity, we rewrite the commuting background \eqref{diag} in terms of the bi-local fields.
First, we introduce the momentum representation 
\bea
\ket{k}=\int d^d k \exp{(ik\cdot x)}\ket{x}.
\label{ketk}
\eea
As usual, we refer to $\{\ket{x}\}$ and $\{\ket{k}\}$ as the basis of the coordinate and momentum representations of $V=\{\phi(x):\mathbb{R}^d \to \mathbb{C}\}$, respectively.
Then we introduce the identification
\bea
\ket{i} \, \in \, \mathbb{C}^N \, \leftrightarrow \, \ket{k=p^i} \, \in \, \{\phi(x):\mathbb{R}^d \to \mathbb{C}\}.
\label{itok}
\eea
By rewriting \eqref{diag} as
\bea
\bra{i}A_\mu^{(0)}\ket{j}=
\left\{
\begin{alignedat}{2}
&p_\mu^i \ \delta_{i,j} \ \ \left(\mu=1 ,\cdots ,d \right)\\
&\ 0 \ \  \left(\mu=d+1 ,\cdots ,10 \right),
\end{alignedat}
\right. 
\qquad \bra{i}\psi_s^{(0)}\ket{j}=0,
\label{diagi}
\eea
it is clear that under the identification \eqref{itok} the solution \eqref{diag} becomes
\bea
\bra{k}A_\mu^{(0)}\ket{k'}=
\left\{
\begin{alignedat}{2}
&k_\mu \, \delta(k-k') \ \left(\mu=1 ,\cdots ,d \right)\\
&\ 0 \qquad\ \left(\mu=d+1 ,\cdots ,10 \right),
\end{alignedat}
\right. 
\qquad \bra{k}\psi_s^{(0)}\ket{k'}=0.
\label{diagk}
\eea
Note that \eqref{diagk} corresponds to the uniform distribution of $p_\mu^i$.
A general distribution of $p_\mu^i$ corresponds to
\bea
\bra{k}A_\mu^{(0)}\ket{k'}=
\left\{
\begin{alignedat}{2}
&f_\mu(k) \, \delta(k-k') \ \left(\mu=1 ,\cdots ,d \right)\\
&\ 0 \qquad \qquad \left(\mu=d+1 ,\cdots ,10 \right),
\end{alignedat}
\right. 
\qquad \bra{k}\psi_s^{(0)}\ket{k'}=0,
\label{diagf}
\eea
where $f_\mu(k)$ is a general function of $k$.

To switch to the coordinate representation, we introduce $\hat{x}^\mu$ and $\hat{p}_\mu$as in ordinary quantum mechanics.
\bea
[\hat{x}^\mu, \, \hat{p}_\nu]=i\,\delta_{\ \nu}^\mu, \quad (\mu,\nu=1.\cdots,d).
\label{xp}
\eea
Then we can express \eqref{diagk} as
\bea
A_\mu^{(0)}=
\left\{
\begin{alignedat}{2}
&\hat{p}_\mu \quad (\mu=1,\cdots, d), \\
&\ 0 \quad \left(\mu=d+1 ,\cdots ,10 \right),
\end{alignedat}
\right. 
\qquad \psi_s^{(0)}=0.
\label{diagop}
\eea
By considering the matrix elements in the coordinate representation, we obtain
\bea
\bra{x}A_\mu^{(0)}\ket{x'}=
\left\{
\begin{alignedat}{2}
&-i \partial_\mu \delta(x-x') \quad (\mu=1,\cdots, d), \\
&\quad 0 \qquad\qquad \left(\mu=d+1 ,\cdots ,10 \right),
\end{alignedat}
\right. 
\quad \bra{x}\psi_s^{(0)}\ket{x'}=0.
\label{unifx}
\eea
The interesting point of this is that the background $A_\mu^{(0)}$ can be regarded as a flat background for the pregeometric action \eqref{bilAct}. 
In fact, the commutator of a bi-local field with $A_\mu^{(0)}$ is the derivative with respect to the center of mass coordinates of the bi-local field as
\bea
\begin{aligned}
\bra{x} [A_\mu^{(0)},\, \phi]\ket{x'} &=\bra{x} [\hat{p}_\mu,\,\phi]\ket{x'} =-i \left(\frac{\partial}{\partial x^\mu}+\frac{\partial}{\partial x'_\mu}\right)\bra{x}\phi\ket{x'} \\
&=-i \frac{\partial}{\partial X^\mu}\phi(X+\frac{\xi}{2}, X-\frac{\xi}{2}),
\label{kinetic}
\end{aligned}
\eea
where $X$ and $\xi$ are the center of mass and the relative coordinates, respectively.
\bea
X=\frac{1}{2}\left(x+x'\right),\quad \xi=x-x'.
\label{Xxi}
\eea

It is useful here to discuss the correspondence with noncommutative spacetime.
As is clear from the representation as a bilocal field, the space of matrices is isomorphic to
\begin{align}
\label{matrix-space-commutative}
    {\rm End}(V) \cong L^2(\R^{2d}).
\end{align}
This can be identified with the (quantized) space of functions on $d+d$-dimensional phase space. 
Viewing matrices as functions on noncommutative space-time is discussed in Section \ref{sec:NC-backgrounds}.

Let us now consider what the general vacuum looks like in terms of bi-local fields if the vacuum is invariant under translations and rotations on $\mathbb{R}^d$. First, from the translational invariance, the one-point Green's functions of the bi-local fields should be functions of the relative coordinates $\xi$:
\bea
\langle \bra{x+a}\phi\ket{x'+a}\rangle_{MM}=\langle \bra{x}\phi\ket{x'}\rangle_{MM}
\, \to \, \langle \bra{x}\phi\ket{x'}\rangle_{MM}=g(x-x')=g(\xi),
\label{tran}
\eea
where $g$ is a function on $\mathbb{R}^d$.
Then considering the rotational invariance in $\mathbb{R}^d$, we have
\bea
\langle \bra{x}A_\mu\ket{x'}\rangle_{MM}=
\left\{
\begin{alignedat}{2}
&-i\,\partial_\mu h(\xi) \quad (\mu=1,\cdots, d), \\
&\quad 0 \qquad\qquad\left(\mu=d+1 ,\cdots ,10 \right),
\end{alignedat}
\right. 
\qquad \langle\bra{x}\psi_s\ket{x'}\rangle_{MM}=0,
\label{genx}
\eea
where $h$ is a rotationally invariant real-valued function on $\mathbb{R}^d$.
Strictly speaking, $\langle A_\mu\rangle$ for $\mu=d+1, \cdots, 10$ could take various values. 
For simplicity, we will assume that the vacuum is symmetric under $SO(d)\times SO(10-d)$ so that they can be set to zero.

In the momentum representation, this becomes
\bea
\langle \bra{k}A_\mu\ket{k'}\rangle_{MM}=
\left\{
\begin{alignedat}{2}
&k_\mu \tilde{h}(k)\delta(k-k') \quad (\mu=1,\cdots, d), \\
&\quad 0 \qquad\qquad\left(\mu=d+1 ,\cdots ,10 \right),
\end{alignedat}
\right. 
\qquad \langle\bra{k}\psi_s\ket{k'}\rangle_{MM}=0,
\label{genk}
\eea
where $\tilde{h}$ is the Fourier transform of $h$.
This is nothing but a rotationally invariant case of \eqref{diagf}.
Therefore, the general distribution of $p_\mu^{(0)\,i}$ in the discrete basis $\{\ket{i}\}$ corresponds to a general translation-invariant vacuum in the coordinate basis $\{\ket{x}\}$.


\section{Graviton as Nambu-Goldstone boson} \label{gra}

From now on, we will consider the Lorentzian signature.
Sometimes it is said that gravitons can be regarded as Nambu-Goldstone bosons \cite{Ph}. 
The argument can be general, but for simplicity, let us consider the Einstein-Hilbert action.
The argument goes as follows:
\begin{enumerate}
\item The basic assumption is that the action is diffeomorphism invariant.
\item If we choose a covariant gauge, such as the harmonic gauge, any constant field configuration, $g_{\mu\nu}(x)=\text{const}$, is a classical vacuum.
Such vacuums are transformed into each other by linear transformations of spacetime, which are special cases of diffeomorphisms:
\bea
g_{\mu\nu} \mapsto g'_{\mu\nu}=M_\mu^{\mu'}M_\nu^{\nu'}g_{\mu'\nu'}, 
\qquad M_\mu^\nu\in GL(d),
\eea
where $d$ is the space-time dimension.
\item The vacuum $g_{\mu\nu}=\eta_{\mu\nu}$ breaks this $GL(d)$ to $SO(d-1,1)$.
\item Therefore, we have degenerate vacuums whose moduli space is $GL(d)/SO(d-1,1)$.
\item Thus, we have Nambu-Goldstone bosons corresponding to
\bea
\text{tangent space of} \, GL(d)/SO(d-1,1)\cong \text{rank 2 symmetric tensor},
\eea
which is nothing but the fluctuations $h_{\mu\nu}$ around the vacuum $\eta_{\mu\nu}$,
\bea
g_{\mu\nu}=\eta_{\mu\nu}+h_{\mu\nu}.
\eea
\item Due to the diffeomorphism invariance, DOF of a vector field can be eliminated. For example, $h_{0\mu}$ can be gauged away, and we have only $h_{ij}$.
\item Taking the residual gauge symmetry into account, only the transverse gravitons remain.
\end{enumerate}

This argument can be generalized to the case of the bi-local field \eqref{bilAct}. 
The most important part of the above discussion is to determine the moduli space of the vacuums. 

To be concrete, we fix the gauge to a covariant gauge such as the Landau type gauge:
\bea
\left[\,\hat{p}_\mu, \,A_\mu\right]=0.
\label{Landau}
\eea
As is shown in \eqref{kinetic}, this is equivalent to
\bea
\frac{\partial}{\partial X^\mu}A_\mu(X+\frac{1}{2}\xi,\,X-\frac{1}{2}\xi)=0,
\label{Landau2}
\eea
where
\bea
A_\mu(x,x')=\bra{x}A_\mu\ket{x'}.
\eea
One good point of this gauge is that it is consistent with the translational invariance of the form
\bea
\frac{\partial}{\partial X^\mu}\langle\bra{x}\phi\ket{x'}\rangle_{MM}=0
\label{trinv}
\eea

As previously discussed, we assume that the vacuum is both translationally and rotationally invariant. 
Then the vacuum expectation values of the bilocal fields are given by a rotationally invariant function of the relative coordinates as \eqref{genx}. 
However, just as $g_{\mu\nu}=\eta_{\mu\nu}$ is not a unique vacuum in general relativity, this is not a unique vacuum. 

As mentioned in section \ref{mat}, we impose the constraint \eqref{unitarityConstraint} on the bi-local fields to guarantee the unitarity.
If this constraint is not imposed, the following arguments \eqref{Lor}-\eqref{mod} hold if VPDs and $SL(d)$ are replaced by diffeomorphisms and $GL(d)$. 
In this case, however, the unitarity is not guaranteed, as we will see in the next section. 
For simplicity, we postpone the consideration of the case with the constraint \eqref{unitarityConstraint} to the end of this section.

Then we can obtain new vacuums by applying Lorentz transformations,
\bea
A_\mu \mapsto A'_\mu=R_\mu^{\,\nu}A_\nu, \qquad R_\mu^{\,\nu}\in SO(d-1,1) \qquad 
\ (\mu,\nu=0,.., d-1),
\label{Lor}
\eea
and linear transformations on $\mathbb{R}^d$ that are special cases of the VPDs, that is,
\bea
X^{\mu} \mapsto X^{\mu\prime} =M^{\mu}_{\nu} X^{\nu}, 
\qquad \xi^{\mu} \mapsto \xi^{\mu\prime} = M^{\mu}_{\nu} \xi^{\nu} \qquad (M \in SL(d)).
\label{SL}
\eea
By applying \eqref{Lor} and \eqref{SL} on \eqref{genx}, we obtain the general vacuum
\bea
\langle \bra{x}A_\mu\ket{x'}\rangle_{MM}=
\left\{
\begin{alignedat}{2}
&-i\,(RM^{-1})_\mu^{\ \nu}\partial_{\nu} h(M\xi) \quad (\mu=0,\cdots, d-1) , \\
&\ \ \qquad \qquad 0 \qquad\qquad\qquad\left(\mu=d ,\cdots ,9 \right).
\end{alignedat}
\right. 
\label{vac}
\eea

Thus $SO(d-1,1)\otimes SL(d)$ acts on the vacuum. 
However, the diagonal $SO(d-1,1)$ subgroup does not change the vacuum. 
That is, if $R=M$, the vacuum does not change because the function $h$ is assumed to be Lorentz invariant.
In the end, the moduli of the vacuum are found to be
\bea
\left(SO(d-1,d)\otimes SL(d)\right)/SO(d-1,1)\cong SL(d).
\label{mod}
\eea
Therefore, we have Nambu-Goldstone bosons corresponding to a $d$-dimensional rank 2 traceless tensor.
Decomposing this tensor into an antisymmetric tensor and a traceless symmetric tensor, we can see that the antisymmetric tensor is the Kalb-Ramond field. The traceless symmetric tensor is similar to the gravity field, but in this case, the 
symmetry is VPD. Using a simple rule of counting the physical degrees of freedom,
\footnote{
Here, the first term, ``$\text{\# of fields}$'', refers to the number of independent component of a symmetric rank-2 tensor, which is $d(d+1)/2$, and the ``dimension of gauge transformation'' is $d-1$ for VPD, so eq.\eqref{dof} equals $(d-2)(d-1)/2 + 1$, which is the sum of the number of physical polarizations of a graviton $(d-2)(d-1)/2$ and $1$ for the dilaton.
}
\bea
\label{dof}
\text{\# of fields}-2\times\text{dimension of gauge transformation},
\eea
we see that there is one more degree of freedom than in general relativity. 
This is because the number of fields is one less, and the dimension of the gauge transformation is one less. 
Eventually, we see that this is the graviton and the dilaton. 
We will examine this in detail in the following three sections.



\section{Low-energy effective theory and the minimal model} \label{low}

To consider the low-energy effective theory for the bilocal fields, we represent bi-local fields as formal power series of differential operators.
\bea
A_\mu=a_\mu^{(0)}(x)+a_\mu^{(1)\,\nu}(x)\partial_\nu+a_\mu^{(2)\,\nu_1\nu_2}(x)\partial_{\nu_1}\partial_{\nu_2}+\cdots.
\label{diffOp}
\eea
The relation between this expression and the bi-local fields can be easily obtained from
\bea
\bra{x}A_\mu\ket{x'}=\int d^d k\bra{x}\left(a_\mu^{(0)}(x)+a_\mu^{(1)\,\nu}(x)\partial_\nu+a_\mu^{(2)\,\nu_1\nu_2}(x)\partial_{\nu_1}\partial_{\nu_2}+\cdots \right)\ket{k}\bra{k}\ket{x'} \nonumber \\
=\int d^d k\left(a_\mu^{(0)}(x)+i\,a_\mu^{(1)\,\nu}(x) k_\nu+i^2\,a_\mu^{(2)\,\nu_1\nu_2}(x)k_{\nu_1}k_{\nu_2}+\cdots \right)\exp(ik\cdot\xi). 
\label{opVsBi2}
\eea
This indicates that the bi-local fields are like generating functions of higher spin fields.

Then the vacuum expectation values of the higher spin fields for the vacuum \eqref{genx} are given by the inverse Fourier transform of \eqref{opVsBi2}.
\bea
\langle a_\mu^{(0)}(x)\rangle_{MM}+i\,\langle a_\mu^{(1)\,\nu}(x)\rangle_{MM} k_\nu+i^2\,\langle a_\mu^{(2)\,\nu_1\nu_2}(x)\rangle_{MM}k_{\nu_1}k_{\nu_2}+\cdots \nonumber \\
=\int d^d \xi \exp(-ik\cdot\xi) \left(-i\,\partial_\mu h(\xi)\right)\ =\,k_\mu \tilde{h}(k) , 
\quad \text{for} \ \mu=0,..,d-1,
\label{vevHigher}
\eea
where $\tilde{h}$ is the Fourier transform of $h$ (see \eqref{genk}).

As we have seen in the previous sections, $h$ represents the distribution of the diagonal elements of $A_\mu$, and at the classical level, an arbitrary $h$ is allowed. 
However, due to quantum fluctuations, the distribution of the diagonal elements is fixed, and the form of $h$ is correspondingly fixed. 
The equation \eqref{vevHigher} means that the vacuum expectation values of various higher-rank tensor fields are determined accordingly.
If these values are uniquely determined, then the excitations around them are generally considered to have mass.

However, as discussed in the previous section, the vacuum is subject to the action of $SO(d-1,1)\otimes SL(d)$. 
If the action does not keep the vacuum invariant, the vacuum is degenerate due to the nontrivial actions. 
Hence, it is natural to expect all higher-rank tensor fields to have mass, except those that appear as Nambu-Goldstone bosons when quantum fluctuations are incorporated.

Let us now discuss the unitarity of the theory, i.e., that all physical states have a positive norm. As discussed around \eqref{Lpt}, if we quench the diagonal elements and consider only planar diagrams, the matrix model is equivalent to the large-$N$ gauge theory. 
Since the latter is a unitary theory, we conclude that the matrix model is also unitary, at least for the off-diagonal elements. 

In more detail, from \eqref{Lpt}, the two-point Green's function of the bilocal field can be written as follows:
\bea
\langle \bra{x_1}A_{\mu_1}\ket{x'_1} \bra{x_2}A_{\mu_2}\ket{x'_2} \rangle_{MM}
=\delta(\xi_1+\xi_2) G_{FT}(X_1, \mu_1, X_2, \mu_2),
\label{2pt}
\eea
where $G_{FT}$ is the 2-point Green's function of the large-$N$ gauge theory with the color factor removed, that is,
\bea
\langle A^{a_1}_{\mu_1}(X_1) A^{a_2}_{\mu_2}(X_2) \rangle_{FT}
=\tr(t^{a_1} t^{a_2}) G_{FT}(X_1, \mu_1, X_2, \mu_2).
\label{2ptFT}
\eea
Here, $X_1, \xi_1$, and $X_2, \xi_2$ are the center of mass and the relative coordinates for 
$x_1, x'_1$ and $x_2, x'_2$, respectively.
From these equations, we see that the degeneracy of the adjoint representation of the large-$N$ gauge theory, $N^2$, corresponds to the infinite number of degrees of freedom that $\xi$ has. This degeneracy is resolved after considering the dynamics of the diagonal elements, as we have discussed. In any case, negative norm states do not appear. Indeed, the inner product
\bea 
\braket{\xi_1| \xi_2}=\delta(\xi_1-\xi_2)
\label{posxi}
\eea
is, at least formally, positive definite on the vector space spanned by $\{\ket{\xi}; \xi \in \mathbb{R}^{d-1,1}\}$.

In general, for massive fields, unitary theories can be created by choosing appropriate coefficients for the terms appearing in the Lagrangian, as in the case of Proca and Fierz-Pauli fields. In this sense, if the matrix model satisfies unitarity as described above, the higher-rank tensor fields should have coefficients well chosen to satisfy unitarity.
On the other hand, for massless fields with spin other than 0 and 1/2, gauge invariance is necessary to satisfy unitarity.

In the present case, the matrix model is invariant under diffeomorphism if we do not consider the constraints as \eqref{unitarityConstraint}, so the low-energy effective theory should also have diffeomorphism invariance as a gauge symmetry.
Among the terms in \eqref{diffOp}, the smaller the number of differential operators, the more important it is at low energy. Therefore, the natural choice for the field describing the low-energy effective theory is one that leaves the first two terms in \eqref{diffOp}:
\bea
A_\mu=a_\mu^{(0)}(x)+a_\mu^{(1)\,\nu}(x)\partial_\nu.
\label{LEdiffOp}
\eea
The first term in the above equation represents the usual abelian gauge field, so we will not consider it specifically here. 
(This field is removed by the constraint~\eqref{unitarityConstraint} -- see Sec.~\ref{constraint}.)
Therefore, we consider
\bea
A_\mu=a_\mu^{\ \nu}(x)\partial_\nu.
\label{LEdiffOp2}
\eea

The equation \eqref{LEdiffOp2} means that the matrix is approximated by a vector field $v^\mu(x)\partial_\mu$. 
The nice thing about it is that the set of such vector fields forms the Lie algebra of the diffeomorphism group of $\mathbb{R}^{d}$. 
Therefore, the equation of motion
\bea
[A_\nu, [A_\mu, A_\nu]]=0
\label{EOMDiffOp}
\eea
can be regarded as an equation on the Lie algebra. 

The diffeomorphism group acts on such $A_\mu$ as gauge transformations.
More explicitly, an infinitesimal diffeomorphism can be written using a vector field $u$ as
\bea
\delta_u A_\mu =[u^\nu \partial_\nu, A_\mu].
\label{infDiff}
\eea
However, as can be seen from the simple number counting, the diffeomorphisms alone are not sufficient to eliminate the negative norm states that $a_\mu ^{\ \nu}$ possesses.
In fact, the negative norm stats contained in $a_\mu ^{\ \nu}$ come from $a_0 ^{\ i}$ and $a_i ^{\ 0}$, which requires $2(d-1)$ gauge degrees of freedom to erase, while the diffeomorphisms have only $d$ degrees of freedom.

However, there is one way to avoid this. It is to impose the condition that the divergence of each component of $A_\mu$ in \eqref{LEdiffOp2} and $u$ in \eqref{infDiff} vanishes:
\bea
\partial_\nu a_\mu^{\ \nu}(x)=0, \label{minimalA}\\
\partial_\nu u^{\nu}(x)=0,
\label{minimalu}
\eea
This is nothing more than restricting the Lie algebra that approximates the matrix from the Lie algebra of the diffeomorphism group to the Lie algebra of the VPD. 
In this case, the advantage of the equations of motion being equations on the Lie algebra is still preserved.
Let us call the model thus obtained the minimal model.
As discussed in the next section, the constraints \eqref{minimalA} and \eqref{minimalu} follow 
naturally at low energy from the constraints \eqref{unitarityConstraint} for the bi-local fields.

Note that this mechanism for unitarity does not work for the higher-spin modes in \eqref{diffOp}. However, as discussed above, these modes are considered to have mass, so such a mechanism is not necessary.


In summary, in the minimal model, the matrices are approximated by the Lie algebra of the VPD, the equations of motion are of the Yang-Mills type, and the gauge symmetry is VPD.
This time, from a simple number counting, it is clear that the negative norm states vanish due to the gauge symmetry. In fact, the negative norm states come from $a_0^{\ i}$, whose degree of freedom $d-1$ is exactly equal to the number of degrees of freedom of the VPD.

In the next section, we propose constraints on the bi-local fields such that the minimal model appears naturally at low energy.
After that, in section \ref{phy}, the physical degrees of freedom of the minimal model are examined in detail.


\section{Constraint for unitarity} \label{constraint}

In the previous section, the minimal model is defined by the constraints \eqref{minimalA} and \eqref{minimalu} imposed on the low-energy effective theory of the original matrix model.
The higher spin components in $A_{\mu}$ are massive, and their gauge transformations are irrelevant.
Nevertheless, it is natural to ask what constraints are imposed with respect to the bilocal fields 
before taking the low-energy limit.
We answer this question in this section.

To find the extension of the constraints \eqref{minimalA} and \eqref{minimalu} to include the higher-spin fields, the crucial requirement is that they are compatible with the gauge transformation:
\begin{equation}
\delta A_{\mu} = i[\Lambda, A_{\mu}].
\label{gauge-transf}
\end{equation}
Including higher-spin fields, we have
\begin{align}
    A_{\mu} &= a_{\mu}(\hat{x}) + a_{\mu}{}^{\nu}(\hat{x}) \hat{p}_{\nu} + a_{\mu}{}^{\nu\lambda}(\hat{x}) \hat{p}_{\nu} \hat{p}_{\lambda} + \cdots,
    \label{A-expand}
    \\
    \Lambda &= u(\hat{x}) + u^{\nu}(\hat{x})\hat{p}_{\nu} + u^{\nu\lambda}(\hat{x}) \hat{p}_{\nu} \hat{p}_{\lambda} + \cdots,
\end{align}
and the gauge transformation \eqref{gauge-transf} implies
\begin{align}
    \delta a_{\mu} &= 
    u^{\nu}\partial_{\nu} a_{\mu} 
    - a_{\mu}{}^{\nu} \partial_{\nu} u,
    \\
    \delta a_{\mu}{}^{\nu} &= u^{\lambda}\partial_{\lambda}a_{\mu}{}^{\nu} - a_{\mu}{}^{\lambda} \partial_{\lambda} u^{\nu} + u^{\nu\lambda}\partial_{\lambda}a_{\mu} - a_{\mu}{}^{\nu\lambda} \partial_{\lambda} u,
    \\
    \vdots & \qquad \quad \vdots
\end{align}
The compatibility of the constraint \eqref{minimalA} with the gauge transformation \eqref{gauge-transf} demands that
\begin{equation}
    \partial_{\nu}(\delta a_{\mu}{}^{\nu}) = 0,
\end{equation}
which implies that
\begin{align}
    \partial_{\lambda}\left(
    u^{\nu\lambda} \partial_{\nu} a_{\mu} - a_{\mu}{}^{\nu\lambda} \partial_{\nu} u
    \right) = 0.
\end{align}
Thus, it is clear that the minimal model constraints \eqref{minimalA}--\eqref{minimalu} have to be modified to conform to the gauge symmetry \eqref{gauge-transf} at general energy scales.

We find a simple, consistent constraint that reduces to eqs.\eqref{minimalA} and \eqref{minimalu} in the low-energy limit: 
\begin{equation}
A_{\mu}|k = 0 \rangle = 0,
\label{matrix-constraint}
\end{equation}
which is indeed satisfied by the background~\eqref{diag}.
Here, $|k = 0 \rangle$ is the state that is identified in the low-energy limit with the eigenstate of the operator $\hat{p}_{\mu}$ with the eigenvalues $k_{\mu} = 0$.
The fermionic field $\Psi$ in the matrix model needs to satisfy the same constraint 
for the sake of supersymmetry.
\begin{equation}
\Psi | k = 0 \rangle  = 0.
\label{matrix-constraint-psi}
\end{equation}

Compatibility of the constraint \eqref{matrix-constraint} with the gauge transformation \eqref{gauge-transf} demands that the gauge transformation parameter $\Lambda$ also satisfies
\begin{equation}
    \Lambda|k = 0 \rangle = 0.
    \label{Lam-constraint}
\end{equation}
Note that \eqref{matrix-constraint}-\eqref{Lam-constraint} are equivalent to \eqref{unitarityConstraint}
because
\begin{equation}
    \ket{k=0}=\int d^d x \ket{x}.
    \label{ket-k=0}
\end{equation}
The effect of these constraints \eqref{matrix-constraint}--\eqref{Lam-constraint} is equivalent to removing the state $\ket{k=0}$ \eqref{ket-k=0} from the representation space $V$.
The state $\ket{k=0}$ is the unique state that is invariant under both the translation and rotation symmetries.
Removing it reduces the gauge symmetry of diffeomorphism to volume-preserving diffeomorphism, thus leading to a unitary quantum theory of gravity as we explained in the previous section.

For a generic matrix configuration including higher-spin fields \eqref{A-expand}, let us redefine the component fields to write it in a manifestly Hermitian way:
\begin{equation}
    A_{\mu} = a_{\mu}(\hat{x}) + \frac{1}{2} (a_{\mu}{}^{\nu}(\hat{x}) \hat{p}_{\nu} + \hat{p}_{\nu} a_{\mu}{}^{\nu}(\hat{x})) + \frac{1}{2} (a_{\mu}{}^{\nu\lambda}(\hat{x})\hat{p}_{\nu}\hat{p}_{\lambda} + \hat{p}_{\nu}\hat{p}_{\lambda} a_{\mu}{}^{\nu\lambda}(\hat{x})) + \cdots,
\end{equation}
where all the fields $a_{\mu}(\hat{x}), a_{\mu}{}^{\nu}(\hat{x}), a_{\mu}{}^{\nu\lambda}(\hat{x}), \cdots$ are real functions of $x$.
The constraint \eqref{matrix-constraint} implies that
\begin{align}
    a_{\mu}(x) - \frac{1}{2} \partial_{\nu}\partial_{\lambda} a_{\mu}{}^{\nu\lambda}(x) + \cdots = 0,
    \label{a-constr-1}
    \\
    \partial_{\nu} a_{\mu}{}^{\nu}(x) - \partial_{\nu}\partial_{\lambda}\partial_{\rho} a_{\mu}{}^{\nu\lambda\rho}(x) + \cdots = 0
    \label{a-constr-2}
\end{align}
for the real and imaginary parts of the constraint, respectively.
Similarly, for the gauge transformation parameter expressed as
\begin{equation}
    \Lambda = u(\hat{x}) + \frac{1}{2} (u^{\nu}(\hat{x}) \hat{p}_{\nu} + \hat{p}_{\nu} u^{\nu}(\hat{x})) + \frac{1}{2} (u^{\nu\lambda}(\hat{x})\hat{p}_{\nu}\hat{p}_{\lambda} + \hat{p}_{\nu}\hat{p}_{\lambda} u^{\nu\lambda}(\hat{x})) + \cdots,
\end{equation}
the constraint \eqref{Lam-constraint} implies
\begin{align}
    u(x) - \frac{1}{2} \partial_{\nu}\partial_{\lambda} u^{\nu\lambda}(x) + \cdots = 0,
    \label{u-constr-1}
    \\
    \partial_{\nu} u^{\nu}(x) - \partial_{\nu}\partial_{\lambda}\partial_{\rho}u^{\nu\lambda\rho}(x) + \cdots = 0.
    \label{u-constr-2}
\end{align}
As higher-spin fields become massive in the low-energy effective theory, the constraints \eqref{a-constr-1}--\eqref{a-constr-2} and \eqref{u-constr-1}--\eqref{u-constr-2} on the lower-spin fields and gauge parameters become
\begin{align}
    a_{\mu} & = 0,
    \label{constr-1}
    \\
    \partial_{\nu} a_{\mu}{}^{\nu} & = 0,
    \label{constr-2}
    \\
    u & = 0,
    \label{constr-3}
    \\
    \partial_{\mu} u^{\mu} & = 0.
    \label{constr-4}
\end{align}
We note that eqs.\eqref{constr-2} and \eqref{constr-4} are identical to eqs.\eqref{minimalA} and \eqref{minimalu}, which we have used to define the minimal model. 
For the sake of the minimal model to be embedded in the original matrix model, what we see here is that we need two more constraints \eqref{constr-1}--\eqref{constr-3} on the minimal model.

The constraints \eqref{matrix-constraint}-\eqref{Lam-constraint} should be upgraded to the original matrix model even for finite $N$.
What is needed is merely the selection of a state ``$|k=0\rangle$'' that plays a special role.


\section{Physical states and difficulties with the minimal model action} \label{phy}

To investigate the physical degrees of freedom that the minimal model has, consider the weak field around the flat space.:
\bea
a_\mu^{\ \nu}(x)=\delta_\mu^{\ \nu}+c_\mu^{\ \nu}.
\label{weakField}
\eea 
At the linearized level, the equation of motion \eqref{EOMDiffOp} and the constraint \eqref{minimalA} become
\bea
\partial_\nu \left(\partial_\mu c_\nu^{\ \lambda}-\partial_\nu c_\mu^{\ \lambda} \right)=0, \qquad \partial_\nu c_\mu^{\ \nu}=0,
\label{linearEOM}
\eea 
and the gauge transformations \eqref{infDiff} and \eqref{minimalu} become
\bea
\delta_u c_\mu^{\ \nu} =\partial_\mu u^\nu, \qquad \partial_\mu u^\mu=0.
\label{linearGT}
\eea 

Using the gauge symmetry, we can fix the gauge to the Lorenz-like gauge:  
\bea
\partial_\mu c_\mu^{\ \nu}=0.
\label{LG}
\eea
Then we have
\bea
\Box \,c_\mu^{\ \nu}=0, \qquad \partial_\nu c_\mu^{\ \nu}=0, \qquad \partial_\mu c_\mu^{\ \nu}=0.
\label{GFixedEOM}
\eea
Thus, we have massless states corresponding to a rank 2 tensor in the transverse $d-2$ dimensions, that is, graviton, Kalb-Ramond, and dilaton, which is the same as the closed string.

More precisely, let us assume that the momentum of the single-particle state is given by
\bea
k=(k^{+}=k, k^{-}=0, k^i=0) 
\label{mom}
\eea
in light cone coordinates.
Then, from the divergence-free constraints \eqref{minimalA} we have
\bea
c_\mu^{\ -}=0. 
\label{m+}
\eea
From the gauge fixing \eqref{LG} we have
\bea
c_+^{\ \nu}=0. 
\label{+n}
\eea
Furthermore, using the residual gauge transformation, 
\bea
\delta \, c_-^{\ \nu}=k\,u^\nu,
\label{resGauge}
\eea
we can set
\bea
 c_-^{\ \nu}=0.
\label{resGauge0}
\eea
Hence, we find that
\bea
c_\mu^{\ \nu} \neq 0 \to \, \mu=i, \ \nu=+ \ \text{or}\, j,
\label{nonzero}
\eea
where $i$ and $j$ stand for the transverse directions.

Thus, the problem of unitarity is to check that the zero-norm states $c_i^{\ +}$ are decoupled from the physical S-matrix. However, this is evident from the fact that the states containing at least one $c_i^{\ +}$ are orthogonal to all state vectors. The important point here is that constraint \eqref{minimalA} is compatible with the equation of motion \eqref{EOMDiffOp} because the VPDs form a group. In other words, requiring both \eqref{minimalA} and \eqref{EOMDiffOp} does not generate a new equation.

Let us conclude this section by examining difficulties with the action of the minimal model. First of all, the naive action 
\bea
S=\frac{1}{4}\tr\left ([A_\mu , A_\nu]^2 \right), \qquad \mbox{where} \quad A_\mu=a_\mu^{\ \nu}\partial_\nu \quad \mbox{and} \quad \partial_\nu a_\mu^{\ \nu}=0,
\label{naiveAction}
\eea
is not finite due to the ultraviolet divergence. 
The trace has
\bea
\tr(u^\mu\partial_\mu \, v^\nu\partial_\nu)=\int d^dx \, d^dy \, u^\mu(x)\,\partial_\mu \delta(x-y) \,v^\nu(x)\,\partial_\nu\delta(y-x)=\infty .
\label{trInf}
\eea

Let us note that the action \eqref{bilAct} itself, written in terms of bilocal fields, is finite. 
It is formally possible to expand the bi-local fields as power series of differential operators as in \eqref{diffOp}, and the equations of motion become coupled differential equations for higher-rank tensor fields with all coefficients being finite and no ultraviolet divergence appearing. 
However, the action is ultraviolet divergent. In other words, expressing bilocal fields as differential operators is a very singular operation in the ultraviolet region.
Thus, when the bi-local fields are expressed or approximated by differential operators, the action is not finite unless the ultraviolet region is appropriately regularized.

The simple idea of regularizing the trace with some ultraviolet cutoff does not work. 
For example, if we regularize the trace as
\bea
\tr_\text{reg}\left(A\right)=\tr\left(A\exp(\epsilon\Box)\right) ,
\label{trReg}
\eea
permutation invariance is violated:
\bea
\tr_\text{reg}\left(A B \right)\neq\tr_\text{reg}\left(B A \right) .
\label{permViol}
\eea
Then the equations of motion no longer have the form \eqref{EOMDiffOp}.

Another idea is to regularize operators themselves, such as
\bea
\left(A\right)_\text{reg}=A\exp\left(\epsilon\Box\right) .
\label{OpReg}
\eea
Then, permutation invariance is satisfied and the equations of motion take the desired form \eqref{EOMDiffOp}. However, due to the regularization, the action becomes a complicated nonlocal functional of the tensor fields.
Indeed, it is a common situation that the equations of motion for the collective coordinates are simple, but the corresponding action is complex and nonlocal \cite{JS}.


\section{Comparison with the
covariant quantum spacetime
approach}
\label{sec:NC-backgrounds}

In this section, we put the formulation considered above in contrast with the approach of
covariant quantum spacetime discussed in Ref.\cite{Sperling:2019xar}. 
To do so, we consider a simple example of the latter approach.
We will see that upon making the time-like matrix $A_0$ slightly off-diagonal, which acts on the "space-like" Hilbert space, the degeneracy of the classical background is reduced, and the action is well-defined and ghost-free for all higher-spin modes analogous to \eqref{diffOp}. Such noncommutative backgrounds admit only finitely many degrees of freedom per unit spacetime volume. On the other hand, because of the explicit violation of Lorentz invariance, the low-energy effective theory is not in perfect agreement with general relativity, and nonlocal corrections appear at a distance.

Here, we consider only 4-dimensional spacetime. Specifically, consider the following background
\bea
A_\mu^{(0)}=
\left\{
\begin{alignedat}{3}
& T_0  \ \left(\mu=0 \right) ,\\
&T_i \ \left(\mu= i =1 ,\cdots ,3 \right), \\
&\ 0 \ \left(\mu=4 ,\cdots ,9 \right).
\end{alignedat}
\right. 
\label{diagk-3}
\eea
Here, $T_0$ and $T_i$ are operators acting on the {\em reduced Hilbert space}
spanned by the space-like momentum states 
\begin{align}
\label{Hilbert-spac-NC}
    \tilde\cH =\langle\ket{k}, \  k \in \R^3 \rangle.
\end{align}
Their explicit forms are given by
\begin{subequations}\label{Tmu}
\begin{align}
&T_0 =  \frac 12 \big(\hat{p}_j \hat{x}^j+ \hat{x}^j \hat{p}_j\big) \ , \\
&T_i=\hat{p}_i \ ,
\end{align}
\end{subequations}
where $\hat{x}^i$ and $\hat{p}_i$ are the standard coordinate and momentum operators acting 
on $\tilde\cH$:
\begin{align}
\label{x-p}
    [\hat{x}^i , \hat{p}_j]=i \delta_j^i \ .
\end{align}
It is important to note that $\hat{p}_i$ and $k_i$ are not to be interpreted as the momentum of a field, but merely a representation of the covariant quantum space algebra.

Note that $T_i$ is diagonal in momentum representation while $T_0$ is slightly off-diagonal.
The important point is that this Hilbert space $\tilde\cH$ consists only of spatial momentum states, and $A_0^{(0)}$ is represented as an operator on $\tilde\cH$. Thus, the noncommutative Hilbert space is much smaller than the commutative Hilbert space.

The operators $T_\mu$ satisfy the commutation relations
\begin{align}
    [T_i,T_j] = 0, \quad 
    [T_0,T_i] = i T_i \ ,
\label{T-CR}
\end{align}
from which we can show that $A_\mu^{(0)}$ given by \eqref{diagk-3} is a solution to the following modified equation of motion. 
\begin{subequations}
\label{modifiedEOM}
\begin{align}
    &\eta^{\mu\nu}[A_\mu^{(0)}, [A_\nu^{(0)},A_0^{(0)}]]= 0, 
\label{EQM0} \\
    &\eta^{\mu\nu}[A_\mu^{(0)}, [A_\nu^{(0)},A_i^{(0)}]]+A_i^{(0)}= 0, 
\label{EQMi}
\end{align}
\end{subequations}
If the quantum effects are approximated by a mass term of the matrix model,
\begin{align}
    \frac{1}{2}\tr(A_i^{\ 2}) , 
\label{massTerm}
\end{align}
$A_\mu^{(0)}$ represents the vacuum state after quantum fluctuations are taken into account.

Next, let us discuss the spacetime represented by $A_\mu^{(0)}$. Assume that operators $Y^\mu$ satisfy the following equations: 
\begin{align}
    [T_\mu,Y^\nu] = f_\mu^\nu(Y),
\label{TY-CR}
\end{align}
where each component of $f_\mu^\nu$ is a function of $Y$'s.
Then, for an operator $\phi(Y)$ given as a function of $Y$s, $[T_\mu, \phi]$ is expressed by the derivatives with respect to $Y$'s of $\phi$. 
\begin{align}
  [T_\mu,\phi(Y)] = f_\mu^\nu(Y)\partial_\nu \phi(Y).
\label{TphiCom}
\end{align}
Here we ignore the operator orderings.
In other words, the commutator with $T_\mu$ can be regarded as a differential operator on the space generated by $Y$'s.

In fact, \eqref{T-CR} can be extended to the full algebra of (minimal) $E(3)$- covariant quantum spacetime $\cM^{3,1}$ \cite{Gass:2025tae} on the minimal doubleton irrep $\tilde \cH \cong \cH_0$ of $SO(4,2)$:
\begin{subequations}
\label{minimal-cov-algebra}
\begin{align}
[T_i,T_j] &= 0, \qquad 
    [T_0,T_i] = i T_i  \\
    [T_\mu, Y^\nu] &= iY^0\delta_\mu^{\nu},
\label{Y0-Yi-CR}  \\
[Y^0,Y^i] &= i T_i,
\\ 
 [Y^i,Y^j] 
&= - i (Y^0)^{-1} (T^i Y^j - T^j Y^i) 
= i \varepsilon^{ijk} L_k
\end{align}
\end{subequations}
with the constraints
\begin{subequations}
\label{minimal-cov-constraints}
\begin{align}
T_i T^i &= Y_0^2 , \label{TT-constraint} \\
T_\mu Y^\mu + Y^\mu T_\mu &= 0 \ ,
\label{T-Y-constraints}
\end{align}
\end{subequations}
where $L_i=\epsilon_{ijk}\hat{x}^j\hat{p}_k$ is the angular momentum operator.
In fact, the explicit forms for $Y^\mu$ in terms of $\hat{x}^i$ and $\hat{p}_i$ are given by
\begin{subequations}\label{Ymu}
\begin{align}
    Y^0 &= \sqrt{\vec{\hat{p}}^2} , \\
    Y^i &= \frac 12(\hat{x}^i Y^0 + Y^0 \hat{x}^i ) ,
\end{align}
\end{subequations}
where $\vec{\hat{p}}^2 = \sum_i \hat{p}_i \hat{p}_i$.
It is straightforward to verify that $T_\mu$ and $Y^\mu$ given by \eqref{Tmu} and \eqref{Ymu} satisfy  the commutation relations \eqref{minimal-cov-algebra}
as well as the constraints \eqref{minimal-cov-constraints}
\footnote{Similar formulas (cf. Appendix G of \cite{Brkic:2024sud}) allow one to obtain general doubleton irreducible representations of $so(4,2)$, leading to non-minimal covariant quantum spacetimes with curvature parameters  $k=0$ or $k=-1$ with slightly modified constraints \cite{Sperling:2019xar,Gass:2025tae}.}.

To visualize the space generated by $Y$'s, let us consider the classical limit where the commutator is replaced by the Poisson bracket. In this limit, \eqref{minimal-cov-constraints} gives a 6-dimensional surface in an 8-dimensional space consisting of $Y_\mu$ and $T^\mu$, while \eqref{Tmu} and \eqref{Ymu} can be regarded as a parameter representation of that 6-dimensional surface by $x^i$ and $p_i$.  More precisely, as can be seen from \eqref{Ymu}, $Y$'s generate a four-dimensional spacetime such that $Y_0 > 0$, and a two-dimensional sphere represented by \eqref{minimal-cov-constraints} is attached at each point on that four-dimensional spacetime. 
It is straightforward to show that the topology of the 6-dimensional surface is identical to $\C P^{1,2}$.
In the following, let us refer to the spacetime generated by $Y$'s as $Y$ spacetime.

Now it is clear from \eqref{TphiCom} and \eqref{Y0-Yi-CR} that $T_\mu$ acts as a derivative operator on the $Y$ spacetime:
\begin{align}
  [T_\mu,.] \sim iy^0\partial_\mu,
\end{align}
where $y^\mu$ stands for the classical limit of $Y_\mu$.
Similarly, if we denote the classical limit of $T_\mu$ as $t_\mu$, we find that
\begin{align}
    \Box := [T^\mu,[T_\mu,.]] \sim -\{ t^\mu,\{t_\mu,\cdot\}\}
\end{align}
 acts as d'Alembertian on the $Y$ spacetime and governs the propagating modes, encoding a $k=0$ FLRW metric $g$ determined by $\Box \sim  \rho^2 \Box_g$\cite{Gass:2025tae}
\footnote{ More general FLRW geometries with curvature parameter $k=0$ are obtained by multiplying $T^0$ with a general function of $Y^0$.
}.

As we have discussed above, due to the constraints 
\eqref{T-Y-constraints}, the matrix algebra
\begin{align}
    {\rm End}(\cH) \sim L^2(\C P^{1,2})
\end{align}
(cf. \eqref{matrix-space-commutative})
can be interpreted as quantized algebra of functions on 
a $S^2$ bundle over the $Y$ spacetime\footnote{This bundle  can be recognized as  twistor space \cite{Steinacker:2023ntw}.} 
 \cite{Gass:2025tae}. 
More explicitly, the full space of modes
is spanned by the higher-spin valued harmonics 
\begin{align}
 \Upsilon_{\Lambda,lm}(y,u)
= \psi_{\Lambda}(y) Y^{lm}(u),
\qquad u^\mu = \frac 1{y_0} t^\mu 
\end{align}
where $Y^{lm}(u)$ are polynomials of order $l$ in the $u^i$,
and 
\begin{align}
 \Box \Upsilon_{\Lambda,lm}(y,u)
&= \big(\Box\psi_{\Lambda}(y)\big) Y^{lm} \ .
\end{align}
This justifies the above interpretation of the background as $S^2$ bundle over the 3+1 dimensional $Y$ spacetime\footnote{The degeneracy of the $Y^{lm}$ is expected to be lifted by quantum effects, giving mass to the higher-spin modes.}.
 The trace becomes the integral over spacetime and averaging over $S^2$:
\begin{align}
    \Tr (\Phi) \sim \int_{\cM^{1,3}} \Omega \, [\phi(y, u)]_0
\end{align}
where $[.]_0$ denotes the integral over the sphere spanned by $u_i$, and $\Omega = \rho_M d^4 y$  with $\rho_M \sim \frac 1{y^0}$ is the (symplectic) volume form on spacetime
Therefore the action is well-defined and does not have the problem of UV divergence as in  \eqref{trInf}
\footnote{A similar background describes a covariant quantum spacetime with curvature parameter $k=-1$ \cite{Sperling:2019xar}.}.

This background differs from commuting backgrounds \eqref{diagk} in several ways.
The space of modes is much smaller than for the commutative background \eqref{matrix-space-commutative}, and admits only finitely many degrees of freedom per unit volume.
Furthermore, the NC background has no flat directions (apart from pure gauge modes and a discrete set of constant higher-spin modes) even classically, and all fluctuations are governed by a 3+1-dimensional FLRW metric
encoded in the matrix d'Alembertian 
$\Box$.
Moreover, the time-like momentum generator $T_0$ is not independent, but determined by the constraint \eqref{T-Y-constraints}. That constraint essentially says that $T_\mu\sim t_\mu$ locally has no time-like component $t_\mu y^\mu = 0$
(consider e.g. the reference point $y^\mu = (y^0,0,0,0)$). This
guarantees that the general (higher-spin-valued) gauge fields $T_\mu \to T_\mu + \tilde A_\mu$ with
\bea
\tilde A_\mu=a_\mu^{(0)}(y)+a_\mu^{(1)\,\nu}(y)u_\nu
+a_\mu^{(2)\,\nu_1\nu_2}(y)u_{\nu_1}u_{\nu_2}+\cdots 
\label{diffOp-NC}
\eea
(cf. \eqref{diffOp}) have no time-like modes, so that no ghosts arise in the Lorentzian action \cite{Steinacker:2019awe}.
 For example, $a_\mu^{(0)}$ plays the role of a Yang-Mills gauge field, while $a_\mu^{(1)\,\nu}$ is a potential for the frame as explained below. 
Nevertheless, physical tensors -- such as the frame and the graviton -- do acquire time-like components due to noncommutativity, as shown below.
The traces \eqref{trInf} are now well-defined as local integrals over spacetime, leading to {\em space-like} tensors upon 
averaging over the local $S^2$ fiber, e.g. 
\begin{align}
    [u_\mu u_\nu]_{0} =: \kappa_{\mu\nu} \ 
\end{align}
The matrix model thus leads to a well-defined (noncommutative but "almost-local") Yang-Mills-type action for $\tilde A_\mu$.
The $E(3)$ symmetry of $k=0$ FLRW spacetime is realized as part of the gauge symmetry 
 via $i[T_i,.]$ and $i[L_i,.]$, which justifies the name "covariant" quantum space. Similarly $e^{i \tau [T_0,.]}$ implements a unitary time-like evolution 
on the space of fluctuations\footnote{This evolution need not coincide with the physical time evolution.}.
 More general gauge transformations realize volume-preserving diffeos $\xi^\mu  = i[\Lambda,y^\mu]_0$ acting on the classical tensorial fields, cf.  \cite{Steinacker:2024unq}. 
This ensures that
gravitons (defined below) remain massless upon quantization\footnote{More precisely, their classical projections transform covariantly. Volume-preserving gauge transformations  include local Lorentz transformations, and suffice to rule out a mass term in the effective action.}, 
just like Yang-Mills gauge fields.

\paragraph{Effective frame and graviton.}

Even though the time-like component of $t_\mu$ vanishes locally as functions, the Poisson brackets $e_\mu = \{t_\mu,.\}$ define a non-degenerate frame including a time-like vector field due to  $\{t_\mu,y^\nu\} = y^0 \delta_\mu^\nu$ \eqref{Y0-Yi-CR}.
More generally, consider a perturbation 
$\tilde A_\mu \sim a_\mu^{\ \nu}(y)u_\nu$
of the above background.
The corresponding frame perturbation
\begin{align}
  \d e_\mu^{\ \nu} \sim \{\tilde A_\mu,y^\nu\}
    = a_\mu^{\ \nu}(y)\{u_\nu,y^\nu\} 
    + [u_\nu\{a_\mu^{\ \nu}(y),y^\nu\}]_0
\end{align}
involves a {\em derivative} of $a_\mu^{\ \nu}$, which is typically the dominant contribution. These derivatives provide the required time-like components.
Hence $a_\mu^{\,\nu}$ should be considered as {\em potential} for the frame $\d e_\mu^{\ \nu}$, which is automatically  divergence-free due to $\partial_\mu(\rho_M \{y^\mu,.\}) = 0$.
Then the graviton 
\begin{align}
    \d g^{\mu\nu} =  \eta^{\a\b} (e_\a^{\ \mu} \d e_\b^{\ \nu}
     + \d e_\a^{\ \mu} e_\b^{\ \nu})
\end{align}
(dropping a conformal factor) transforms as a rank-2 tensor under volume-preserving diffeomorphisms.
We see here that the gravitons arise in a somewhat different way.

This observation
implies that the classical (Yang-Mills-type) matrix model action \eqref{IKKT} becomes {\em non-local} in terms of the frame $e_\mu^{\ \nu}$. While it does admit Ricci-flat linearized metric fluctuation \cite{Sperling:2019xar}, it cannot be interpreted as the Einstein-Hilbert action. The E-H action arises in the quantum effective action at one loop, which is UV finite and approximately local on the noncommutative background with finite density of states,
due to maximal supersymmetry  \cite{Steinacker:2023myp}. This 1-loop action includes additional terms for the dilaton and (gravitational) axion.

As a consequence, the one-loop effective gravitational action is a combination of the E-H action with the classical matrix action, which includes non-local terms in the frame. This combination leads to an interesting modification of GR at large distances, which will be discussed in detail elsewhere.

\section{Summary} \label{sum}


The mechanism by which spacetime emerges in the IIB matrix model is not yet fully understood.
In particular, there are several different possibilities as to what kind of variables related to spacetime the matrix represents. 
For example, it is possible that the matrices represent the coordinates of spacetime, or the momenta of fields propagating through spacetime, or the coordinates of non-commutative spacetime.
If the fluctuations of the matrices are fully taken into account, these scenarios may be equivalent, but this is not yet fully understood. 

The fundamental question is how general relativity arises as a low-energy effective theory in these scenarios.
The first scenario is where the matrix corresponds to the coordinates of space-time as a function on the worldsheet of string theory. In this scenario, general relativity should appear as the low-energy effective theory of closed strings.
The main purpose of this paper is to examine how general relativity appears in the second scenario.
After that, we have considered the third scenario in comparison with the second scenario.

As for the second scenario, where the matrices represent momenta, the basic assumption is that the diagonal elements of the matrices are $d$-dimensionally continuously distributed under appropriate gauge fixings, such as the Landau gauge, and further that the distribution is Lorentz invariant.

The following points should be emphasized. 
At the level of classical solutions, the distribution of diagonal elements is completely arbitrary, since any simultaneous diagonal matrices are classical solutions. 
On the other hand, when quantum fluctuations are incorporated, non-trivial interactions between diagonal elements are induced. 
The assumption here is that the distribution should be almost uniquely determined as a result of this interaction. 
By nearly unique, we mean unique except for the degeneracy guaranteed by symmetry.
The degeneracy remaining due to the symmetry is nothing but the moduli of the vacuum. 
The massless modes that represent their fluctuations are gravity, dilaton, and the Kalb-Ramond field.

Another point to note is that the positive definiteness (unitarity) of the bilocal fields is self-evident
 in the planar sector, as shown in \eqref{2pt} and \eqref{posxi}. 
 This point is very different from the case where the bilocal fields are expressed as power series of differential operators. 
 In that case, the time component of each tensor field has a negative norm. 
 However, as we have seen, the expansion of bilocal fields by differential operators is highly singular in the ultraviolet region.
In particular, there is no clear correspondence between the inner product as a bilocal field and the inner product of the tensor field when expanded by differential operators.

In summary, when matrices are represented as bilocal fields, the action of the IIB matrix model and the inner product of states do not involve ultraviolet divergence. 
On the other hand, when attempting to express the theory using tensor fields by expanding the matrices as differential operators, both the action and the inner product of states are ultraviolet divergent. 

We have analyzed the structure of the vacuum and the behavior of fluctuations around it at low energies by expressing the matrices in terms of bilocal fields. 
Although the bi-local fields could be non-local, in the low-energy limit, they are reduced to local fields consisting of a finite number of massless fields and an infinite number of massive fields.
Massive fields do not require special consideration to ensure unitarity, but in the case of massless fields, some gauge symmetries or constraints are necessary to eliminate negative norm states.
We have proposed simple constraints on the bi-local fields and proved that they indeed work.
As a result, we find that general relativity emerges spontaneously around the spacetime generated by the IIB matrix model,
assuming the mass gap for higher-spin fields.


As a concrete example of the third scenario, where the matrices represent a non-commutative spacetime, we have considered a (minimally) noncommutative background. 
Here, the time-like matrix $A_0^{(0)}$ acts on a reduced Hilbert space spanned by the space-like eigenstates.
The basic difference between the second and the third scenarios can be stated as follows: A Lorentz-invariant commutative vacuum requires four independent generators leading to bi-local fields on $\R^4$, while the noncommutative vacuum is based on three independent generators leading to bi-local fields on $\R^3$. 
This makes the unitarity manifest and solves the difficulty of expanding bi-local fields in terms of differential operators.
However, Lorentz invariance is no longer manifest,
and non-local modifications to general relativity appear in the infrared region.


\section*{Acknowledgments}

We would like to thank Antal Jevicki and Wei-Hsiang Shao for useful discussions. 
H.S. would like to acknowledge related collaborations with Christian Gass and Alessandro Manta.
P. M. H. is supported in part by the National Science and Technology Council, R.O.C. (NSTC 113-2112-M-002 -040 -MY2), and by National Taiwan University.
H.K. is partially supported by JSPS (Grants-in-Aid for Scientific Research Grants No. 20K03970), 
and by National Taiwan University.
H.K. also thanks Prof. Shin-Nan Yang and his family for their kind support through the Chin-Yu chair professorship.
The work of H.S. is supported by the Austrian Science Fund (FWF) grant P36479.

\end{document}